\documentclass[12pt,preprint]{aastex}

\newcommand{\diaz}{N$_2$H$^+$}
\newcommand{\ceo}{C$^{18}$O}

\slugcomment{Version 4.1 - Aug 18, 2009}

\shorttitle{Molecular Line Observations of Infrared Dark Clouds II}
\shortauthors{Gibson et al.}

\begin{document}


\title{Molecular Line Observations of Infrared Dark Clouds II:  Physical Conditions}


\author{David Gibson\altaffilmark{1}, Ren\'e Plume\altaffilmark{1,3}}

\author{Edwin Bergin\altaffilmark{2} , Sarah Ragan\altaffilmark{2}}

\and

\author{Natalie Evans\altaffilmark{1}}

\altaffiltext{1}{Department of Physics and Astronomy, University of Calgary,
2500 University Drive NW, Calgary, Alberta, T2N 1N4 Canada}
\altaffiltext{2}{Department of Astronomy, University of Michigan, 500 Church Street, Ann Arbor, MI 48109, United States}
\altaffiltext{3}{Max Planck Institute for Astronomy, K$\ddot{o}$ningstuhl 17, 69117 Heidelberg, Germany}




\begin{abstract} 

Using a source selection biased towards high mass star forming regions, we used a Large Velocity Gradient (LVG) code to calculate the H$_2$ densities and CS column densities for a sample of Midcourse Space Experiment (MSX) 8$\mu$m infrared dark cores. 
 Our average H$_2$ density and CS column density were 1.14 x 10$^{6}$ cm$^{-3}$ and 1.21 x 10$^{13}$ cm$^{-2}$ respectively.  In addition, we have calculated the Jeans mass and Virial mass for each core to get a better understanding of their gravitational stability.  We found that core masses calculated from observations of N$_2$H$^+$ J = 1$\rightarrow$0 and C$^{18}$O J = 1$\rightarrow$0 by Ragan et al. 2006 (Paper 1) were sufficient for collapse, though most regions are likely to form protoclusters.   We have explored the star-forming properties of the molecular gas within our sample and find some diversity which extends the range of infrared dark clouds from very the massive clouds that will create large clusters, to clouds that are similar to some of our local counterparts (e.g. Serpens, Ophiuchus).
\end{abstract}


\section{Introduction}

Over the past decade Infrared Dark Clouds (IRDC) have become to be recognized as the sites of clustered star formation in our galaxy (Rathborne et al. 2006; Ragan, Bergin, \& Gutermuth 2009; Butler \& Tan 2009).   IRDCs are important not only in understanding the earliest stage of the birth of stellar clusters, which is the dominant mode of galactic star formation (Lada \& Lada   2003), but also as the primary sites of massive star formation.   Massive star formation plays an integral role in the evolution of galaxies as they are responsible for energizing the interstellar medium (McKee 1986), producing the heavy elements (Dopita 1991), and have been suggested to regulate the rate of star formation (McKee \& Tan 2003).  Despite their importance, a variety of observational and theoretical issues have made massive star formation a difficult topic to address.    In particular the greater distances towards massive star forming regions, faster evolutionary timescales, and source confusion (e.g. Garay \& Lizano 1999) make it difficult to isolate the earliest stages of massive star formation.   Since IRDCs represent the earliest stages of star cluster formation they provide fertile ground for understanding the beginnings of high and low mass stellar birth (Menten, Pillai, \& Wyrowski 2005, Bergin \& Tafalla 2007).

There have been several studies of IRDCs that have characterized the masses and column density structure in these objects with the ultimate aim of understanding clustered and massive star formation  (Sridharan et al. 2005; Pillai et al. 2006; Rathborne et al. 2006; Ragan et al. 2006; Du \& Yang 2008).   What is clear is that these objects are massive, with masses in excess of 100 M$_{\odot}$ which is comparable to the envelopes around some massive protostars.   The velocity linewidths of 1--3 km s$^{-1}$ are also characteristically below that of molecular gas surrounding massive stars, which is believed to relate to the fact that the IR dark clouds are in an earlier phase of star formation.  In addition they are often associated with the signposts of star formation including masers 
(Beuther et al. 2002;   Wang et al. 2006), outflows (Beuther \& Sridharan 2007), and embedded infrared point sources (Beuther \& Steinacker 2007, Chambers et al. 2009; Ragan et al. 2009).

Since IRDCs have only been recently recognized there is a basic lack of large scale information on various physical characteristics, such as the temperature and densities, that can be reliably determined via well known molecular-line techniques.      Pillai et al. (2006; 2007) and Sridharan et al. (2002) used the NH$_3$ inversion transitions to explore temperature variations within a sample of 32  and 40 IRDCs respectively and found that gas temperatures are typically 20~K.   Direct determination of the densities from H$_2$CO (Carey et al. 1998), CH$_3$OH (Leurini et al. 2007), and CS (Rathborne et al 2008) are available for about 10--20 objects with densities found to be in excess of $10^{5}$ cm$^{-3}$.
It is the aim of this work to provide some of this basic information to further illuminate the general characteristics of IRDCs in the context of massive star formation.  In particular we will use multi-line observations of CS to estimate the gas density for a larger sample of 41 infrared dark clouds isolated from the MSX database by Ragan et al. (2006).   Using this information we will explore the diversity amongst a modest portion of the IRDC population and grossly characterize the stability of these objects against gravitational collapse.

 In sections 2 and 3 we will present our observations and place the sources into different categories of molecular emission morphologies based on observations from Ragan et al. (2006) (hereafter Paper 1) and new CS observations.  We will then present the results of large velocity gradient (LVG) analysis of these cores to constrain their H$_2$ densities and CS column densities.  In section 4, we will present the evidence that confirms several of our selected cores are actually massive star forming regions, and assign relative chemical ages to these cores using a chemical evolution model.  

\section{Observations}
\subsection{Source Selection}

There have been numerous studies of the initial conditions of high mass star formation using a variety of criteria with which to select sources.  For example, some have used IRAS colors and/or water masers as signposts of star formation (e.g. Molinari et al. 1996, Plume et al. 1992, 1997; Knez et al. 2002; Mueller et al. 2002; Sridharan et al. 2002).  Others have surveyed a sample of infrared dark cores (IRDCs) (e.g. Redman et al. 2002; Sridharan et al. 2005; Rathborne et al. 2006; Pillai et al. 2006) and some have selected their sample from the mapping of molecular cores (e.g. Wu et al. 2000; Garay et al. 2004).  All of these studies, however, have both benefits and drawbacks.  For example, the water maser studies could not disentangle massive young stellar objects (YSO) that have already formed from those in the initial stages of collapse.  On the other hand, the IRDC studies cannot confirm that the clouds mapped will indeed form massive YSOs. 

In Paper 1 we presented a biased sample of potential massive protostellar cores from the 8.8 $\mu$m images of the Midcourse Space Experiment (MSX; Price et al. 1996) satellite.  This sample is biased in that, to maximize the possibility that our sample actually contains massive prestellar objects, we have constrained our search to IR dark cores that are located close to Ultra Compact HII (UCHII) regions on the plane of the sky.  Given that massive stars tend to form in clusters (Lada, Strom \& Myers 1993), current massive star forming regions are excellent places to look for potential massive protostellar cores.  In particular, we looked for cores that were IR dark at 8 $\mu$m (suggesting either starless cores or deeply embedded YSOs) that were near UCHII regions (signposts of very young massive stars (Churchwell 2002)).  

Using this criteria, we identified 114 objects of which only a small fraction (15$\%$) have known associations with radio sources or masers.  Of these, 41 were selected due to their compactness (R$_{core}$ $<$ 2.1 pc), and relatively high opacity ($\tau$ $\stackrel{>}{\sim}$ 0.4 at 8.8 $\mu$m).  In this paper, we present new observations of the 41 IRDCs presented in Paper 1.

\subsection{Molecular Line Observations}

In Paper 1, we presented observations of C$^{18}$O J = 1 $\rightarrow$ 0 ($\nu$ = 109.782 GHz), CS J = 2 $\rightarrow$ 1 ($\nu$ = 97.981 GHz) and N$_2$H$^+$ J = 1 $\rightarrow$ 0 ($\nu$ = 93.173 GHz) which were observed at the 14m FCRAO in February 2002, May 2002 and December 2002.  Using the 16 element focal plane array receiver SEQUOIA, each region was mapped with 17 x 17 point maps spanning 400$''$ x 400$''$, typically to a 1$\sigma$ rms noise level of $\sim$ 0.05 - 0.1K.  We used the Narrow Band Correlator backend, which was configured to a velocity resolution of $\sim$ 0.13 km s$^{-1}$.  System temperatures were typically 200 - 300K.  Main beam efficiencies ($\eta_{mb}$) were assumed to be 0.48, taken from standard FCRAO values.  The FCRAO beamsize at these frequencies is $\Theta_{FWHM} \sim$ 45$''$.  The emission from CO  J = 1 $\rightarrow$ 0 ($\nu$ = 115.271 GHz) was also observed at the 14m FCRAO over the same time period as the other observations using the same configurations, except the $^{12}$CO observations were performed in poorer weather conditions (T$_{sys}$ $\sim$  750K).  Though not used in Paper 1, these CO observations have been used in this publication (See Table 1 for results).  

In this paper, we focus on new  CS observations.  Emission from CS J = 3 $\rightarrow$ 2 ($\nu$ = 146.969 GHz) and CS J = 5 $\rightarrow$ 4 ($\nu$ = 244.936 GHz) was observed at the central position of the MSX cores (constrained using the peak absorption in the MSX Band A image)  in October 2002, February 2003, March 2003, June 2003, and April 2004.  However, in a few sources  the MSX core was not co-located with the intensity peak of  the previously obtained CS, CO and  C$^{18}$O molecular line maps.  In these cases, the new CS observations were taken on the center of the MSX core, the position of which is listed as an offset ($\Delta \alpha$, $\Delta \delta$) from the center of the previously obtained molecular line maps.  

The CS J = 3 $\rightarrow$ 2 was observed at the Kitt Peak 12m telescope using Filter bank spectrometers FB12 and FB22, as well as the Millimeter Auto Correlator (MAC).  These instruments were configured to a velocity resolution of $\sim$ 0.2 km s$^{-1}$, 0.51 km s$^{-1}$, and 0.05 km s$^{-1}$ respectively, and typically yield a 1$\sigma$ rms noise level of $\sim$ 0.05 - 0.1K.  System temperatures were typically 300 - 400K, with a beamsize of $\Theta_{FWHM}$ = 45$''$ and a main beam efficiency of $\eta_{mb}$ = 0.48.  CS J = 5 $\rightarrow$ 4 was observed at the 10m Heinrich Hertz Submillimeter Telescope (HHSMT) using acousto-optical spectrometers B and C, and the Chirp Transform Spectrometer, configured to a velocity resolution of 0.5876 km s$^{-1}$,  0.1481 km s$^{-1}$, and 0.057 km s$^{-1}$ respectively.  The central position of each MSX core was observed with a 37$''$ beam ($\eta_{MB}$ = 0.82), with a typical rms noise level of $\sim$ 0.05 - 0.5K.  System temperatures were typically 300-400K, however the observations made in June 2003 were  performed in poorer weather conditions raising the temperature (T$_{sys}$ $\sim$ 800K).  Results of these observations can be found in Tables 2 and 3.  

The emission from CS J = 7 $\rightarrow$ 6 ($\nu$ = 342.882 GHz) was observed at the 10.4m Caltech Submillimeter Observatory (CSO) in July 2002, using the 280 - 400 GHz receiver and the AOS backend configured to a velocity resolution of $\sim$ 0.042 km s$^{-1}$.  The central position of each MSX core was observed with a 24.6$''$ beam ($\eta_{MB}$ = 0.75).  Typical system temperatures were 600 - 650K.  Results of these observations can be found in Table 3.  

All data are presented in units of T$_A$$^*$ (Kutner \& Ulich 1981).  However, to compare our observations to model predictions and to calculate physical conditions, we must include the main beam coupling efficiency ($\eta_{mb}$) such that T$_{mb}$ = $T_A^*/\eta_{mb}$.  All data were calibrated with the standard chopper wheel method.

\section{Results}
\subsection{Molecular Emission}

Tables 2 and 3 show the results of Gaussian fits to the observed CS transitions.  The tables consists of the following entries (labeled as columns (1) through (11)): (1) core name, (2) \& (3) the position offsets ($\Delta\alpha$ and $\Delta\delta$ respectively) , (4) \& (8) the antenna temperature (T$_a$$^*$), (5) \& (9) the full width half maximum (FWHM), (6) \& (10) the source v$_{lsr}$, and (7) \& (11) the integrated intensity ($\int T_a^*dv$).  Average line widths at the core's central position are 2.85 km s$^{-1}$ for CS J = 2 $\rightarrow$ 1, 2.68 km s$^{-1}$ for CS J = 3 $\rightarrow$ 2, 3.09 km s$^{-1}$ for CS J = 5 $\rightarrow$ 4, and 2.93 km s$^{-1}$ for CS J = 7 $\rightarrow$ 6.

In Paper 1, we presented a catalog of maps displaying the spatial distributions of C$^{18}$O J = 1 $\rightarrow$ 0, CS J = 2 $\rightarrow$ 1, and N$_2$H$^+$ J = 1 $\rightarrow$ 0 for each source.  Based on the spatial distributions, we are able to separate our sources into 4 broad categories.  CO is not included in this categorization process since it is not as useful a discriminant as is C$^{18}$O.  Table 4 shows each core's morphological category based upon the following criteria: 

\begin{itemize}
\item Category 1 - CS J = 2 $\rightarrow$ 1 and/or C$^{18}$O J = 1 $\rightarrow$ 0 emission detectable and centralized on the MSX dark core, but no detectable N$_2$H$^+$ J = 1 $\rightarrow$ 0.  Five cores fall in this category:  G06.26-0.51, G09.16+0.06, G14.33-0.57, G24.16+0.08, and G37.44+0.14.

\item Category 2 - Centralized CS J = 2 $\rightarrow$ 1 and/or C$^{18}$O J = 1 $\rightarrow$ 0 emission, with  detectable N$_2$H$^+$  J = 1 $\rightarrow$ 0 emission.  Seventeen cores fall in this category:  G09.21-0.22, G09.28-0.15, G09.86-0.04, G10.99-0.09, G12.22+0.14, G12.50-0.22, G19.37-0.03, G23.37-0.29, G24.05-0.22, G25.99-0.06, G30.98-0.15, G32.01+0.05, G34.63-1.03, G34.74-0.12, G34.78-0.80, G35.20-0.72, and G37.89-0.15

\item Category 3 - Strong N$_2$H$^+$ J = 1 $\rightarrow$ 0 emission, with reduced CS J = 2 $\rightarrow$ 1 and C$^{18}$O J = 1 $\rightarrow$ 0 abundances based on their average column densities (see Table 4 for complete list of column densities).  Five cores fall in this category:  G05.85-0.23, G19.40-0.01, G23.48-0.53, G30.89+0.14, and G31.02-0.12

\item Category 4  - Little to no emission was detected in any of the molecular transitions observed.  Fourteen cores fall in this category:  G09.88-0.11, G10.59-0.31, G10.70-0.33 , G19.28-0.39, G30.14-0.07, G30.49-0.39, G30.53-0.27, G33.82-0.22, G43.64-0.82, G43.78+0.05, G50.07+0.06, G53.88-0.18, G75.75+0.75, and G76.38+0.63.  These cores are not included in Table 4.  
\end{itemize}

\begin{figure}[T]
\begin {center}
\includegraphics[width=4in, angle=-90]{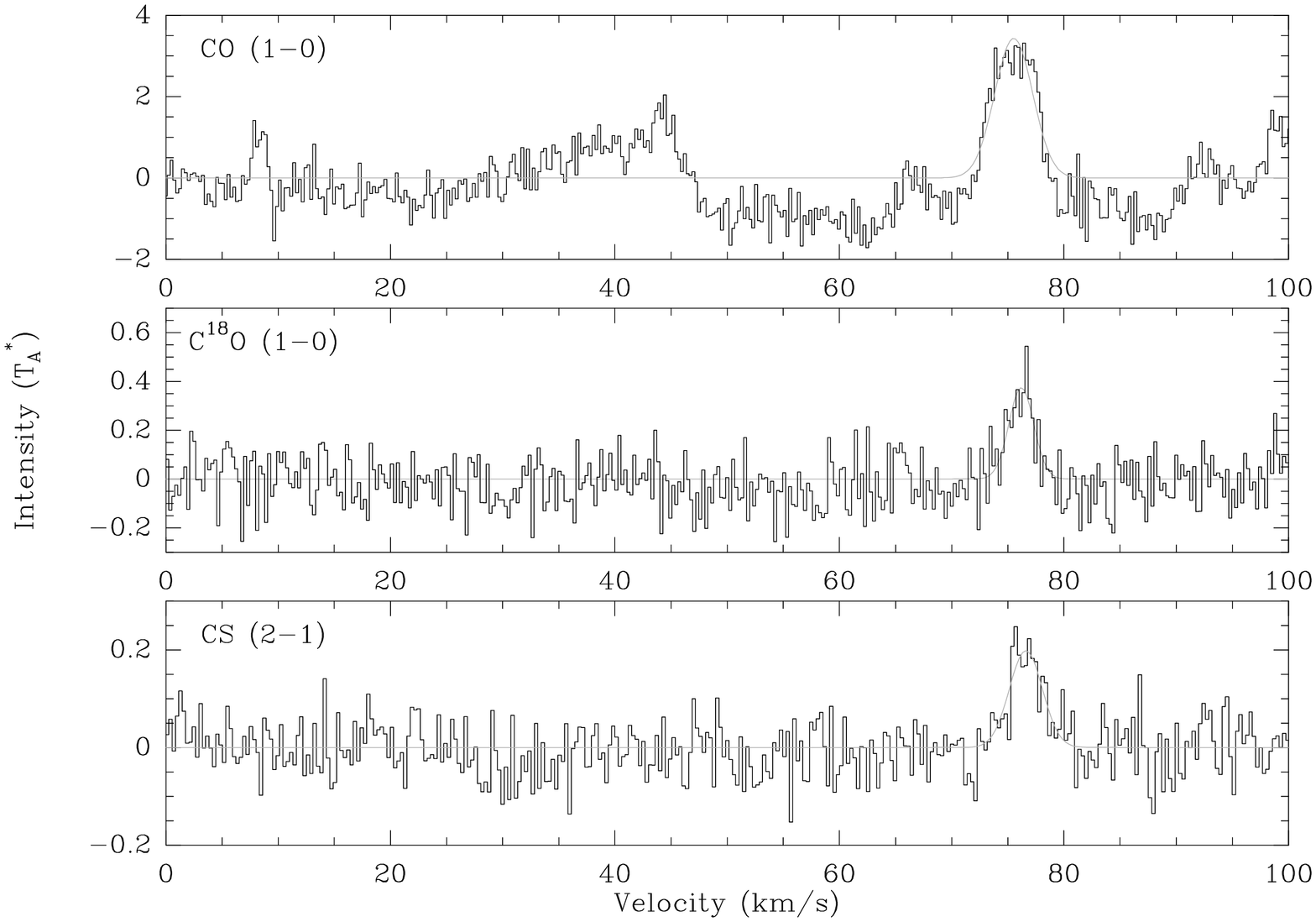}
\caption{Determining the Velocity of Core G31.02-0.12.  Shown are three molecules that were observed for the given core.  As can be seen, though there are multiple velocity components in the CO data, the true velocity of the MSX core can be determined through additional observations of another optically thin and/or high density tracer.}
\label{velocity}
\end{center}
\end{figure}

In several cases, there were examples of multiple velocity components (e.g G31.02-0.12, see Figure \ref{velocity}) likely due to line of sight clouds.  In these cases the velocity of each core was determined by analyzing the emission from multiple molecular lines for coincidence.  We have made the assumption that CS, as a tracer of dense gas, preferentially selects the presumably dense core rather than the low density line-of-sight emission.  Figure \ref{velocity} shows an example of this analysis.  

In some other cases, molecular line emission peaks off the central position of the core (G10.59-0.31, 35.20-0.72 for example).  This behavior could be caused by either a nearby object heating the surface of the cloud, internal heating from an embedded protostar, or abundance gradients causing the molecular emission to peak off the cold core center.  The existence of possible embedded protostars has been explored in a sub-sample of these objects by Ragan et al. (2009).  However in these cases, we still believe the molecular emission is associated with the MSX cores and have been categorized accordingly.  Alternatively, from the morphologies of G19.40-0.01 in Figure 1 from Paper 1, for example, it is quite obvious that although the IR dark core is located in the center of the image, the molecular emission is centered on a secondary source below it.  Other examples of this can be found in G9.21-0.22 and G32.01+0.05.  However, in these cases, despite the molecular emission appearing to be centered on a secondary source, they have been classified as Category 2 objects as they still exhibit strong emission of all tracer molecules on the core center.  

\subsection{Physical Conditions}

In Paper 1, we presented simple LTE calculations of the \ceo, \diaz, and CS column densities and used these to extract the average H$_2$ column density and cloud mass.  Table 4, columns 3 and 4 reproduces the results of the \ceo\ and \diaz\ column density calculations, with C$^{18}$O converted to CO assuming an abundance ratio of N$_{CO}$ = 500N$_{C^{18}O}$.  Here we present a more detailed analysis of the emission from multiple CS transitions to more accurately determine the gas density.  We have presented only the 27  cores that fall in categories 1 to 3, omitting the 14 category 4 cores.

\subsubsection{Temperature}

Observations of CO J = 1 $\rightarrow$ 0 and C$^{18}$O J = 1 $\rightarrow$ 0 were used to estimate the kinetic temperature of each region.  First, it was assumed that CO J = 1 $\rightarrow$ 0 is optically thick and in LTE.  These are reasonable assumptions given the low critical density and excitation temperature of CO J = 1 $\rightarrow$ 0, and the large abundance of CO J = 1 $\rightarrow$ 0 in the ISM (10$^{-4}$ with respect to H$_2$).  Under these assumptions T$_{mb}$ $\sim$ T$_k$, meaning that the observed main beam temperature (T$_{mb}$) is a good approximation of the kinetic temperature of the region.  CO and C$^{18}$O emission lines were generally fit by a single Gaussian (self absorption effects were negligible).  Next, we used our optically thin C$^{18}$O J = 1 $\rightarrow$ 0 as a check for these kinetic temperatures assuming an abundance ratio of N$_{CO}$ = 500N$_{C^{18}O}$.  The results were within a few percent of the temperatures derived  from the CO data, with the average kinetic temperature found to be $\sim$ 8K, ranging between 4K and 16K.  These CO temperatures are lower than the 15 - 20K typically found in previous observations of similar regions, often through the use of NH$_3$ line ratios (e.g. Garay et al. 2004; Sridharan et al. 2005; Pillai et al. 2006).  Such low CO temperatures could be  caused if the regions are clumpy, with substructure smaller than the beam (i.e. beam filling factors $<$ 1). Therefore, given the unknown filling factor, we have assumed a canonical kinetic temperature of 15K to use in all our subsequent analysis.

\subsubsection{LVG Analysis}

To determine the H$_2$ densities and CS column densities, we used a large velocity gradient (LVG) code (as used by Plume et al. 1997) to solve the coupled equations of statistical equilibrium and radiative transfer.  Using the assumptions of constant density and temperature with a uniformly filled beam, we created a 20 $\times$ 20 grid in column density per velocity interval - density space with our H$_2$ densities ranging from 10$^4$ to 10$^8$ cm$^{-3}$ and our CS column density per velocity interval ranging from 10$^{11}$ to 10$^{18}$ cm$^{-2}$ (km s$^{-1}$)$^{-1}$.  These ranges are meant to straddle the expected values of density and column density found in other high mass star forming regions (e.g. Plume et al. 1997).  Each density-column density combination produced a series of model line intensities.  These model line intensities were then compared to our CS J = 2 $\rightarrow$ 1, J = 3 $\rightarrow$ 2,  J = 5 $\rightarrow$ 4, and J = 7 $\rightarrow$ 6 observations via a reduced $\chi ^2$ minimization routine.  Thus, the $\chi ^2$ routine finds which of the 400 models best fits our data, providing us with the density and column density of the core. 

In some sources, the higher J transitions were too weak to be detected: e.g. 6 of the category 1 to 3 sources did not have detectable CS J = 5 $\rightarrow$ 4 nor 7 $\rightarrow$6 emission, and an additional 9 sources were detected in everything but CS J = 7 $\rightarrow$ 6.  For the sources without detectable J = 5 $\rightarrow$ 4 and 7 $\rightarrow$6 emission, we adopted the $1\sigma$ rms noise level of the J = 5 $\rightarrow$ 4 observation to assign an upper limit to the line intensity of this transition and assumed the line width and doppler velocity to be the same for all transitions.  This provided us with at least three data points (with the third point being an upper limit in 6 of the sources) to which we could fit the LVG models, since a fit to only two data points would not provide valid results. The resulting densities and column densities can be found in Table 4, columns 5 and 6.

In an additional 7 of the category 1 to 3 sources, we detect only J = 2 $\rightarrow$ 1 data.  In these cases we calculated the CS column density under the assumptions that the transition was optically thin and in LTE.  To verify that this assumption is acceptable, we used the full LVG model to calculate the line opacities for a simulated cloud with N$_{CS}$ = 1.21 x 10$^{13}$ cm$^{-2}$ and n$_{H_2}$ = 1.14 x 10$^{6}$ cm$^{-3}$ (the typical values determined from LVG fitting to all cores for which we had 3 or 4 CS transitions).  The resulting opacity from the model was 0.1, which supports our optically thin assumption.  A similar verification was also done  for sources where the higher J transitions were detected.  Column densities for these sources calculated assuming that the CS J = 2 $\rightarrow$ 1 transition was optically thin and in LTE were within a few percent of those calculated using the full LVG analysis.

\section{Discussion}

\subsection{Are These Site of Massive Star Formation?}

To determine which cores from our selection are potential sites of massive star formation, we used a variety of observed properties as evidence, specifically line widths, and column densities, along with a comparison to previous studies of star forming regions, both low mass and high mass.

\subsubsection{Line Widths}

Crapsi et al. (2005) and Caselli et al. (2002) observed mean line widths of $\sim$ 0.33 km s$^{-1}$ and 0.50 km s$^{-1}$ for N$_2$H$^+$ in low mass star forming regions.  On the other hand, Pillai et al. (2006) and Sridharan et al (2005) observed mean line widths of $\sim$ 2 km s$^{-1}$ and 2.1 km s$^{-1}$ respectively in regions they concluded to be in the early stages of massive star formation.  In addition, Pillai et al. (2007) observed line widths of $\sim$ 2.7 km s$^{-1}$ in potential prestellar clumps in proximity to UCHII regions.  Plume et al. (1997) and Shirley et al. (2003), observed mean line widths of 4.2 and 5.6 km s$^{-1}$ for CS in high mass star forming regions.  Their regions, however, are active regions of star formation suggesting that the broader line widths are caused by turbulence injected by outflows from young stars.

The observed mean CS line width in both the category 2 and 3 sources is 3.0 km s$^{-1}$, whereas in the category 1 cores the means CS line width is 2.5 km s$^{-1}$.  These are consistent with the results of Pillai et al. (2007).  Based on these results, we conclude that our regions have more non-thermal support than low mass star forming regions, and are broad enough to be possible high mass star forming regions that are not currently active (see also discussion in Paper 1).   The fact that our lines are narrower than those found by Plume et al. (1997) and Shirley et al. (2003) suggests that our cores are more quiescent and possibly at an earlier evolutionary stage.  One caveat is that line widths of $>$ 2.0 km s$^{-1}$ can occur in low mass star forming regions with active outflows (Rudolph et al. 2001; Tafalla et al. 2004).  However, our spectra show no evidence of line wing emission which would indicate active outflows.

\subsubsection{Column Densities and Densities}

The average CS column density in Table 4 is 1.21 x 10$^{13}$ cm$^{-2}$.  This is approximately an order of magnitude higher than the CS column densities seen in the sample of low mass cores by Snell, Langer, \& Frerking (1982) and Zhou et al. (1989).  Our average N$_2$H$^+$ column density of 7.24 x 10$^{13}$ cm$^{-2}$ (Paper 1) is also an order of magnitude higher than that found by both Caselli et al. (2002) and Crapsi et al. (2005) which were both studies of low mass star forming regions.  See Figure 2 for the distribution of CS column densities.  

Beuther et al. (2002) found N$_{CS}$ to be $\sim$ 10$^{12}$ - 10$^{16}$ cm$^{-2}$ in massive star forming regions in the early stages of evolution.  In active high mass star forming regions, Lada et al. (1997) found an average N$_{CS} = 3.98 \times 10^{13}$ cm$^{-2}$, comparable to our results.  Plume et al. (1997), however, found N$_{CS}$ = 2.63 x 10$^{14}$ cm$^{-2}$ which is an order of magnitude higher than the CS column densities derived for our sample of cores.  This is likely a real difference in column density between their maser sources and our IRDCs, and not just a modeling effect.  For example, when Plume et al. (1997) modeled their same sources, but ignored the CS J = 7 $\rightarrow$6 detections, they still found $<N_{CS}> = 2.29 \times 10^{14}$ cm$^{-2}$.   In contrast, in their sources without any CS J = 7 $\rightarrow$ 6 detections, they found N$_{CS}= 3.71 \times 10^{13}$ cm$^{-2}$, which is comparable to our results. Thus, the reason that we do not detect CS J = 7$\rightarrow$6 in most of our cores is most probably a column density effect.


Figure 3 shows a histogram of the density distributions in our cores.  Our average H$_2$ density from LVG analysis is 1.1 $\times 10^{6}$ cm$^{-3}$.  In comparison, Beuther et al. (2002) found an average H$_2$ density of $\sim$ 10$^6$ cm$^{-3}$ in young massive star forming regions.  Plume et al. (1997) found an average of  8.5 $\times 10^{5}$ cm$^{-3}$ and Lada et al. (1997) found an average of 6.3 $\times 10^{5}$ cm$^{-3}$ in active high mass star forming regions.  In contrast, Zhou et al. (1989) found that 2.0 $\times 10^{5}$ cm$^{-3}$ in regions forming low mass stars.  It is interesting to note that the densities of our cores are similar to those in the water maser survey of Plume et al. (1997) while our column densities are an order of magnitude lower.  This may be a beam filling factor effect caused by ``clumps'' that are smaller than the beam.  In this case, the ``clumps'' in our sample of IRDCs would have the same density as those in the water maser survey of Plume et al. (1997), but they would be smaller or fewer in number and, thus, fill less of the beam resulting in a lower beam-averaged column density.  

Thus the large linewidths, high densities, and column densities in our sample of MSX cores support the idea that these cores are, indeed, potential sites for high mass star formation.

\subsection{Are These Objects Young?}

An understanding of interstellar chemistry combined with a model based on the object's temperature and density can provide a rough estimate to a core's age using the relative abundances of C$^{18}$O, CS, and N$_2$H$^+$.  We used a chemical evolution code developed at the University of Calgary, which, provided a series of input parameters (n, T, UV field, etc.), solves the coupled differential equations which determine the abundances of each species for each time step.  Thus, the code allows us to track the relative abundances of any species as a function of time.  Our code incorporates gas-phase reactions, gas-grain interactions (depletion and desorption) through the treatment described by Hasegawa \& Herbst (1993), and Bergin et al. (1995), using both cosmic ray desorption and thermal evaporation as destruction mechanisms.  The code utilizes the RATE99 (Le Teuff et al. 2000) reaction rate database which provides 4100 reactions connecting 400 species.  Standard relative atomic abundances were used (see Table 5 which also sets the initial abundances of species at time t = 0) with a set UV field strength of 1 Habing (reasonable for protostellar cores at a temperature of 15K in proximity to an UCHII region (Churchwell 2002)), a grain albedo of 0.6, and an A$_v$ $>$ 100 implying the UV field is shielded in the interior.  For the sake of computational speed, we created a scaled down version of this database, with 270 species.  When tested against the full database, we saw little difference between the results.  

To model our data, we ran this code using the density and temperature which emulates an average core (n = 1.14 x 10$^{6}$ cm$^{-3}$, T = 15 K).  Figure 4a shows the results for CO, CS, and N$_2$H$^+$.  It is important to note that small changes in density and temperature reflecting the range of physical conditions seen in our cores produced no significant difference in the model.  In the early stages of evolution (t $<$ 10$^2$ years), CO and CS abundances increase rapidly to their maximum level, yet there is very little (if any) N$_2$H$^+$ present.  This is mainly due to the difference in reaction rates  and formation pathways which allow CO  and CS to form rapidly.  Whereas CO, as the principle destroyer of N$_2$H$^+$ (Aikawa et al. 2001), hinders the increase in abundance of N$_2$H$^+$.  Beyond $\sim$ 10$^5$ years CO and CS begin to deplete quite readily onto dust grains which allows the N$_2$H$^+$ abundance to increase.  In addition, although N$_2$H$^+$ can be destroyed in the gas phase through dissociative recombination with electrons, this process simply produces N$_2$ which does not easily deplete onto dust grains but instead rapidly reacts with H$_3$$^+$ to reform N$_2$H$^+$ (Nakano \& Umebayashi 1986).  

Although Figure 4a presents molecular abundance relative to the total Hydrogen abundance (H + H$_2$), we do not have direct information on the Hydrogen abundance in these cores for us to directly compare our observed abundances to the model.  However, by using the ratios of N$_{CO}$/N$_{CS}$, N$_{CO}$/N$_{N_2H^+}$, and N$_{CS}$/N$_{N_2H^+}$ , we eliminate the H abundance allowing us to compare our observations to the model and assign a relative chemical age.  Nevertheless in Figure 4a, we attempt to provide absolute abundances based on approximate H$_2$ column densities from Paper 1 by convolving our MSX images to the resolution of our FCRAO data and using the simple relation $\tau_{\lambda}$ $=$ $\sigma_{\lambda}$ $\cdot$ N(H$_2$), where $\tau_{\lambda}$ is the dust opacity, $\sigma_{\lambda}$ is the dust extinction cross section (2.3$\times 10^{-23}$ cm$^{2}$ at 8.8 $\mu$m; Indebetouw et al. 2005), and N(H$_2$) is the column density of molecular hydrogen.  

Figure 4b shows the column density ratio plots on which we have placed the positions of each core.  The average CS column density and relative abundance of category 1 cores are 7.1 x 10$^{12}$ cm$^{-2}$ and 1.22 x 10$^{-10}$ respectively, however with no detectable \diaz abundance above our detection limits (X/(H + H$_2$) $\sim$ 2.0 x 10$^{-13}$) these objects must be chemically young (t $<$ 10$^2$ years).   An alternate interpretation of these data is that there is an embedded source in the vicinity which could release CO and CS from grain mantles and destroy N$_2$H$^+$ (e.g. Lee et al. 2001).  In this instance the core would be more evolved but would appear to be chemically younger since  the chemical clock has been reset.   Careful analysis of the Spitzer images by Ragan et al. (2009), however, reveals no 24 $\mu$m sources coincident with the category 1 sources G06.26-0.51 and G06.16+0.06 indicating that they may, indeed, be extremely young objects devoid of any star formation activity (unless any existing 24 $\mu$m sources are extremely deeply embedded).  G37.44+0.14, however, appears to contain a Class I protostar and, our cursory inspection of the Spitzer data for G24.16+0.08, also indicates the presence of 24 $\mu$m source coincident with the 8 $\mu$m dark image (there are no MIPS data for G14.33-0.57).  Thus, it is possible that these objects are older but have had their chemical clocks reset due to heating by the embedded sources.

Category 2 cores, which have a  detection of all three molecular tracers, have an apparent ambiguity in their relative ages as their CO and CS abundances can indicate an extremely chemically young (t $<$ 10 years) or more chemically evolved core (t $>$ 10$^{4.5}$ years).  However, the absolute abundance of \diaz\ at the younger ages would be well below our detection limit and inspection of Figure 4b shows that the abundance ratios also preclude the possibility of these young ages.  Thus, we conclude that these cores are indeed more chemically evolved.  In category 3 cores, we see reduced abundances of CS (X/(H + H$_2$) $\sim$ 1.04 x 10$^{-10}$)(compared to the average relative CS abundance from our full sample) coupled with a larger column density of N$_2$H$^+$ (X/(H + H$_2$) $\sim$ 3.43 x 10$^{-11}$), which could be evidence of depletion.  In low mass star forming regions depletion is dominant in the time period before protostellar formation (indicative of a more chemically evolved region), and a similar trend is expected to be seen for high mass star forming regions.  Looking at the observed column density ratios versus  those from our model (see Figure 4b),  we assign an age range of 10$^{4.6}$ $<$ t $<$ 10$^{5.1}$ years for category 3 objects.  This suggests these sources are more chemically evolved, and could be at the onset of star formation.

It is evident from Figure 4b that, despite our different definitions, category 2 and 3 objects appear to have similar  chemical ages.  The presence of an embedded star would certainly allow CO to evaporate off dust grains giving it  the appearance of a chemically younger object (category 2) despite being more chemically evolved (category 3).  However the similarity in ages is more likely a result of the steepness of the abundances profiles in log space.  Around 10$^5$ years, large differences in CS or CO column densities translate into fairly small changes in time.  Since we have limited observational sensitivity, the corresponding  chemical ages of category 2 and 3 cores will necessarily be similar.  

The absolute ages of the cores, however, is also subject to model assumptions, such as  our depletion and desorption rates.  Bergin, Langer, \& Goldsmith (1995)  found that desorption of species from grain mantles has a strong dependance on its surface binding energy, which is highly uncertain and depends upon which species is the dominant component of the grain mantle.  We tested this effect on our results by adjusting our binding energies and cosmic ray desorption flux by as little as 10 percent, which resulted in the chemical ages changing by as much as a factor of a few.  
 Therefore given our observational errors and uncertainties in our modeling, we cannot provide an absolute ages for category 2 and 3 cores but suggest that they are somewhere between 10$^{4.5}$ $<$ t $<$ 10$^{5.5}$ years .  We also cannot provide a firm age for category 1 cores, but their complete lack  of N$_2$H$^+$ emission is a strong argument in favor of their extreme youth.

\subsection{A Diversity of IRDCs}

In Paper 1, we presented core masses based upon observations of C$^{18}$O J = 1 $\rightarrow$ 0 and N$_2$H$^+$ J = 1 $\rightarrow$ 0, with kinematic distances calculated using the line center velocity and the Milky Way rotation curve model of Fich, Blitz, \& Stark (1989).  However, in the early stages of core evolution mass can be primarily accounted for by C$^{18}$O, whereas in later stages when it is depleted, the sum of this depleted C$^{18}$O and the newly formed N$_2$H$^+$ is likely a better representation of the overall mass. Therefore, in Table 6, column 6 of this paper, we provide the sum of these masses.  As a means of comparison to these molecular line based core masses, it is useful to calculate the virial mass of the cores.  We used the relation:
\begin{equation}
M_{Virial} = \frac{5}{3}\frac{RV_{rms}^2}{G}
\end{equation}
where  R is the radius of the cloud, G the gravitational constant and V$_{rms}$ = 3$^{\frac{1}{2}}$$\Delta V/2.35$ where $\Delta$V is the average FWHM linewidth of the CS observations.   Since the virial mass is only dependent on the radius (and not the cube of the radius) it is much less sensitive to errors in the assumed distance to the cloud.  Virial masses of the cores can be found in Table 6, column 5.

Comparison of the molecular line based masses to the virial masses may provide information about the structure and state of the cores.  The average M$_{Virial}$/M$_{Molecular}$ is 0.78 suggesting that the clouds are roughly in Virial equilibrium.   Note, however,  that in only 4 cases M$_{Virial}$ $>$ M$_{mol}$ whereas, in the rest, M$_{Virial}$ $<$ M$_{mol}$.  These discrepancies could be caused by a number of factors.  For example, if the clouds are clumpy with beam filling factors less than 1, the observed intensity will be lower than it would be for a uniformly filled beam, resulting in lower column density values and, therefore, artificially lowered values of M$_{mol}$.
Indeed in a sub-sample of our cores, Ragan et al. 2009, find clear evidence for significant sub-structure on scales below
the beam size of all our molecular observations. 
  Other possibilities are the existence of streaming motions adding to the line width of the observations resulting in artificially increased values of M$_{vir}$.  Alternately, the cores may not actually be in Virial Equilibrium, suggesting  that they are not gravitationally bound and may, in fact, be  transient objects (Ballesteros-Paredes 2006).  Finally, our acceptance of standard abundance ratios may be n\"aive, as many studies have shown this value to vary with location (e.g. Wall 2007).  

Comparison of the Jeans mass to the mass calculated from our molecular observations allows us to estimate if our cores are gravitationally unstable and whether or not the core has the potential to form a single star or a cluster.  Using a canonical temperature of 15K and the density results from the LVG model, we used the following reduced form of the Jeans mass relation:
\begin{equation}
M_{Jeans} = (\frac{5kT}{G\mu m_H})^{\frac{3}{2}}(\frac{3}{4\pi \rho_0})^{\frac{1}{2}} M_{\odot}
\label{jeans}
\end{equation}
where $\rho_0$ is the density of the region, and $\mu$ is the mean mass per particle (2.2).  Jeans masses of the cores can be found in Table 5, column 4.  Cores with only one CS transition observed or detected do not have a Jeans mass since, in these cases, LVG modeling to obtain their densities could not be undertaken.

Table 6, shows that almost all sources we observed have masses substantially larger than the Jeans mass, in many cases 2 - 3 orders of magnitude higher in value, suggesting that these sources will not form individual protostars but rather protoclusters.  This is not unusual since star formation is known to be strongly clustered (Lada et al. 1997; Bonnell \& Davies 1998) and IRDCs are clearly the precursors to stellar clusters (e.g. Rathborne et al. 2006, Ragan et al. 2009).

Ragan et al. 2009 performed a detailed investigation of the association of mid-IR absorbing cloud with young stars as identified by the Spitzer bands in a sub-sample of sources in this paper.   For all sources  they found that most of the cloud mass is unassociated with deeply embedded young protostars.  However,  all of the cores in their smaller Spitzer  sub-sample  have young stellar objects present in the near vicinity, just  beyond the confines of the mid-IR dark cloud.   Thus, to the limit of the Spitzer observations, which is to below a solar mass at the cloud surface, and larger masses for deeply embedded sources, the bulk of the molecular material appears pre-stellar.   In particular, G05.85-0.23 and G37.89-0.15 had no embedded sources at 24 $\mu$m within the boundaries of the cores, with a maximum of 3 embedded sources found along the edges of G34.63-1.03 and G37.44-0.14.  Labeled a category 1 object, the embedded sources within G37.44-0.14 likely evaporated CO from the dust grains and destroyed N$_2$H$^+$, which would cause an error in our chemical model interpretation.  

In this regard there are two cores that stand out as somewhat different than the rest of the sample: G37.44-0.14, and G37.89-0.15.  These cores appear to have molecular masses sufficient for the formation of massive stars (Sridharan et al. 2005 found masses to be a few x 10$^2$ - 10$^3$ M$_{\odot}$ for high mass pre-protostellar objects).  In addition their virial and jeans masses suggest these objects are likely gravitational bound and massive enough to collapse.  However unlike other sources that also pass these criteria, their cloud masses are only 8 and 3 times larger than their Jeans masses and their densities are about an order of magnitude below the average value.   Thus, these cores will likely form less massive clusters or possibly even single, massive stars.  Although, in the latter case, this single ``isolated'' massive star would likely be forming in the presence of other young stellar objects in the context of a larger cluster.  For example, in G37.44, there are other 8$\mu$m dark regions in the MSX image presented in Paper  1 which may all be related to a larger complex.
These results suggest a diversity in the type of stellar systems that can be formed within infrared dark clouds - from very massive clouds that will create large clusters, to smaller clouds that are analogs of some of the local star-forming regions like Serpens and Ophiuchus.

\section{Conclusions}

We have identified 27 possible protostellar objects through the observation of molecular tracers C$^{18}$O, CS, and N$_2$H$^+$.  By calculating their column densities and densities, we determined their evolutionary states based on a comparison to a chemical evolution model.  In Paper 1 the primary focus was on the morphologies of the molecular emission of these tracers, with rough estimates for their sizes and densities. In this paper, we used the following evidence to identify these sources as potential sites of massive star formation:
\begin{enumerate}
\item Our average CS column density (1.21 x 10$^{13}$ cm$^{-2}$) and H$_2$ density (1.14 x 10$^{6}$ cm$^{-3}$) is higher then those found in low mass star forming regions (e.g. Snell et al. 1982) and comparable to those found in high mass star forming regions (e.g. Plume et al. 1997).
\item Our average CS line widths ($\Delta v$ $\sim$ 2.0 - 3.0 km s$^{-1}$) are comparable to high mass star forming regions (e.g. Pillai et al. 2007), and larger than low mass star forming regions (e.g. Crapsi et al. 2005).
\item Our core masses are larger than their Jeans masses suggesting they are of sufficient mass to collapse.  In addition, M$_{Virial}$/M$_{Molecular}$ = 0.78 which suggests that these regions are likely gravitationally bound.  
\item By comparing our molecular abundances to a chemical model, we found our sources to be chemically young (10$^{4.5}$ $<$ t $<$ 10$^{5.5}$ years) for category 2 and 3, t $<$ 10$^2$ years for category 1) suggesting that these regions may not have yet formed massive protostars.  
\end{enumerate}

In addition, we have explored the generic star-forming properties within the sample and find some diversity in the types of stellar systems that can be formed within IRDCs thereby extending the range of infrared dark clouds from the very massive clouds that will create large clusters to clouds that are analogs of some of the local star-forming regions.
Multiwavelength studies of the internal structure of these cores using the Spitzer Space Telescope (Ragan et al. 2009) along with high resolution molecular observations to further constrain their temperature and density structure will help to provide a more accurate picture of these regions as well as the nature of their embedded stellar populations.

The authors would like to thank the anonymous referee who provided many useful comments and who helped to significantly improve this paper.

\clearpage

\begin{figure}[t]
\begin {center}
\includegraphics[width=4in, angle=-90]{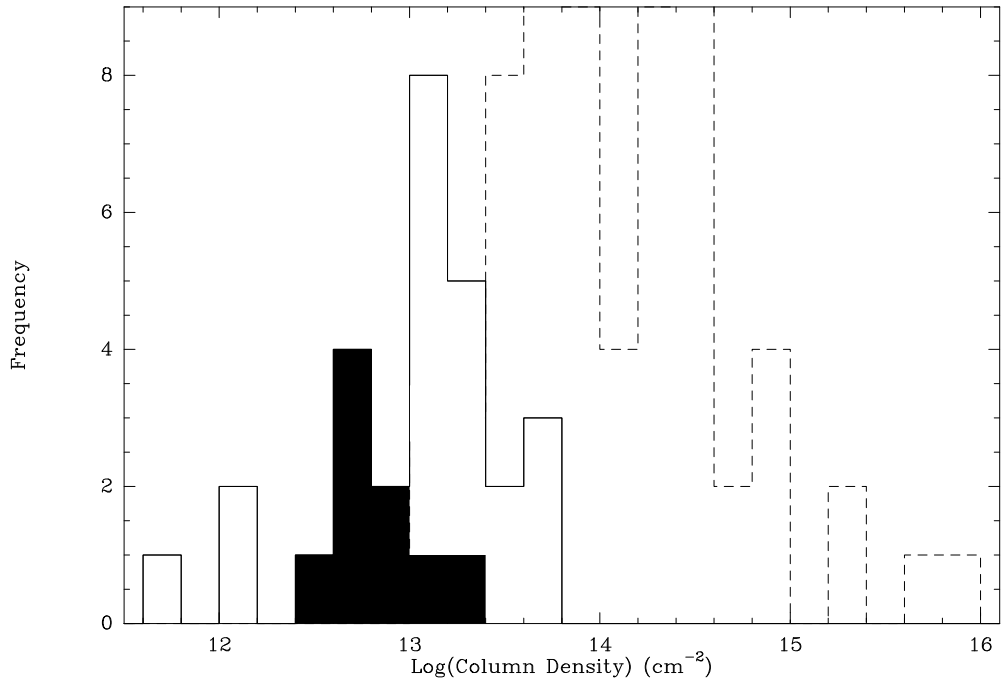}
\caption{Solid line - Histogram of the CS column densities found using LVG analysis.  The average column density is 1.21 x 10$^{13}$ cm$^{-2}$ with a 1$\sigma$ standard deviation of 0.49.  Dashed line - Results from Plume et al. (1997) with an average CS column density of 1.07 x 10$^{14}$ cm$^{-2}$ (peak of histogram off scale).  Filled area - Results from Snell et al. (1982), with an average density of 7.53 x 10$^{12}$ cm$^{-2}$.}
\label{density}
\end{center}
\end{figure}

\begin{figure}[t]
\begin {center}

\includegraphics[width=4in, angle=-90]{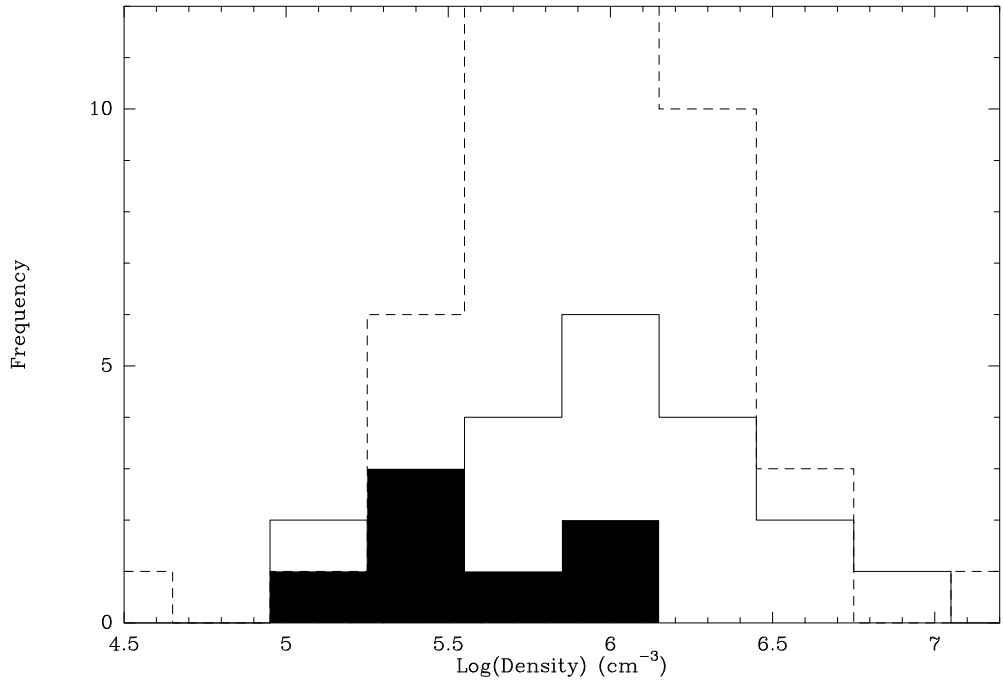}
\caption{Solid line - Histogram of the H$_2$ densities found using LVG analysis.  The average density is 1.14 x 10$^6$ cm$^{-3}$ with a 1$\sigma$ standard deviation of 0.53.  Dashed line - Results from Plume et al. (1997), with an average density of 8.51 x 10$^6$ cm$^{-3}$ (peak of histogram off scale).  Filled area - Results from Lada et al. (1997), with an average density of 6.31 x 10$^{5}$ cm$^{-3}$}
\label{density}
\end{center}
\end{figure}

\begin{figure}[t]
\begin{center}
\includegraphics[width=4in, angle=-90]{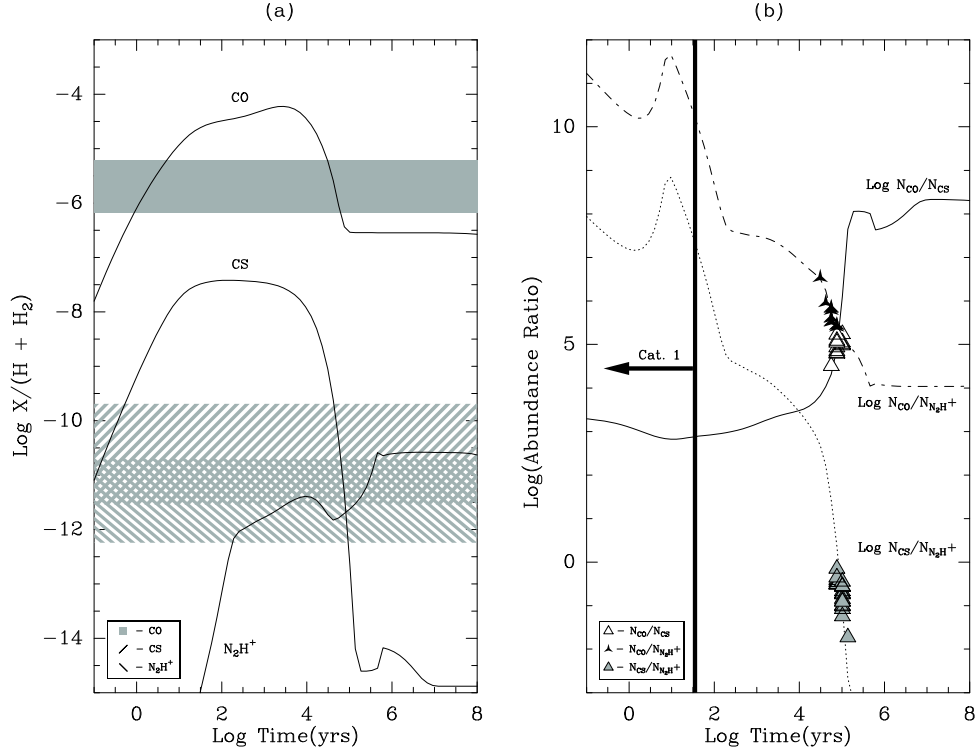}
\caption{Standard Chemical Evolution Model for Sample Cores. 
(a):  Chemical model generated by our chemical evolution code, showing the relationship between the abundances of CO, CS, and N$_2$H$^+$  (T = 15 K, n = 1.14 x 10$^6$ cm$^{-3}$) showing chemical equilibrium being reached at t $\sim$ 10$^6$ years.  The range of abundances observed for each species is indicated by the listed fill pattern.  (b):  Abundances ratios of N$_{CO}$/N$_{CS}$, N$_{CO}$/N$_{N_2H^+}$, N$_{CS}$/N$_{N_2H^+}$ from our model, showing an inverse relations between CO, CS, and N$_2$H$^+$.   The column density ratios of category 2 and 3 cores are presented as markers on the model lines.  Category 1 cores have an upper limit of t $<$ 10$^2$ years.}
\label{chemmodel}
\end{center}
\end{figure}

\clearpage

\begin{deluxetable}{lrrrcccrccc} 
\tablecolumns{7} 
\tablewidth{0pc} 
\tablecaption{Observational Parameters at Center of MSX Core for CO J = 1 $\rightarrow$ 0} 
\tablehead{ 
\colhead{Core} & \colhead{$\Delta\alpha$}   & \colhead{$\Delta\delta$ }    & \colhead{$T_a^*$} & 
\colhead{$\Delta v$}    & \colhead{$v_{lsr}$}   & \colhead{$\int T_a^*dv$}\\
& \colhead{($''$)} & \colhead{($''$)} & \colhead{(K)} & \colhead{(km s$^{-1}$)} & \colhead{(km s$^{-1}$)} & \colhead{(K km s$^{-1}$)}\\
\colhead{(1)} & \colhead{(2)} & \colhead{(3)} & \colhead{(4)} & \colhead{(5)} & \colhead{(6)} & \colhead{(7)}}
\startdata
G05.85-0.23	&	25	&	-25	&	2.25	&	3.25	&	16.85	&	7.76	\\
G06.26-0.51	&	0	&	0	&	3.55	&	4.99	&	21.89	&	18.85	\\
G09.16+0.06	&	0	&	0	&	2.98	&	3.14	&	31.89	&	9.96	\\
G09.21-0.22	&	0	&	0	&	\nodata	&	\nodata	&	\nodata	&	\nodata	\\
G09.28-0.15	&	0	&	0	&	\nodata	&	\nodata	&	\nodata	&	\nodata	\\
G09.86-0.04	&	0	&	0	&	\nodata	&	\nodata	&	\nodata	&	\nodata	\\
G09.88-0.11	&	0	&	0	&	\nodata	&	\nodata	&	\nodata	&	\nodata	\\
G10.59-0.31	&	0	&	0	&	\nodata	&	\nodata	&	\nodata	&	\nodata	\\
G10.70-0.33 	&	0	&	0	&	\nodata	&	\nodata	&	\nodata	&	\nodata	\\
G10.99-0.09 	&	0	&	0	&	1.11	&	4.28	&	29.20	&	4.60	\\
G12.22+0.14	&	0	&	0	&	\nodata	&	\nodata	&	\nodata	&	\nodata	\\
G12.50-0.22	&	0	&	-25	&	0.88	&	2.51	&	35.82	&	3.65	\\
G14.33-0.57	&	0	&	0	&	4.60	&	8.09	&	21.27	&	39.61	\\
G19.28-0.39	&	0	&	0	&	\nodata	&	\nodata	&	\nodata	&	\nodata	\\
G19.37-0.03	&	0	&	0	&	\nodata	&	\nodata	&	\nodata	&	\nodata	\\
G19.40-0.01	&	0	&	0	&	\nodata	&	\nodata	&	\nodata	&	\nodata	\\
G23.37-0.29	&	0	&	0	&	2.69	&	7.16	&	76.57	&	20.47	\\
G23.48-0.53	&	0	&	0	&	\nodata	&	\nodata	&	\nodata	&	\nodata	\\
G24.05-0.22	&	25	&	0	&	3.32	&	5.84	&	81.52	&	20.61	\\
G24.16+0.08	&	0	&	0	&	2.10	&	9.64	&	109.55	&	21.56	\\
G25.99-0.06	&	25	&	25	&	\nodata	&	\nodata	&	\nodata	&	\nodata	\\
G30.14-0.07	&	0	&	0	&	\nodata	&	\nodata	&	\nodata	&	\nodata	\\
G30.49-0.39	&	0	&	0	&	\nodata	&	\nodata	&	\nodata	&	\nodata	\\
G30.53-0.27	&	0	&	0	&	\nodata	&	\nodata	&	\nodata	&	\nodata	\\
G30.89+0.14	&	0	&	0	&	\nodata	&	\nodata	&	\nodata	&	\nodata	\\
G30.98-0.15	&	-25	&	0	&	\nodata	&	\nodata	&	\nodata	&	\nodata	\\
G31.02-0.12	&	-25	&	0	&	3.52	&	4.03	&	75.98	&	15.09	\\
G32.01+0.05	&	0	&	0	&	7.10	&	7.78	&	96.25	&	58.80	\\
G33.82-0.22	&	0	&	0	&	$<$ 0.46	&	\nodata	&	\nodata	&	\nodata	\\
G34.63-1.03	&	0	&	0	&	1.19	&	2.35	&	11.58	&	2.97	\\
G34.74-0.12	&	-25	&	0	&	2.25	&	9.78	&	77.35	&	23.43	\\
G34.78-0.80	&	0	&	0	&	3.43	&	4.56	&	43.30	&	42.32	\\
G35.20-0.72	&	0	&	0	&	3.14	&	6.28	&	32.86	&	21.00	\\
G37.44+0.14	&	0	&	0	&	7.67	&	2.51	&	39.96	&	20.46	\\
G37.89-0.15	&	0	&	-25	&	2.18	&	1.25	&	13.08	&	2.91	\\
G43.64-0.82	&	0	&	0	&	\nodata	&	\nodata	&	\nodata	&	\nodata	\\
G43.78+0.05	&	0	&	0	&	\nodata	&	\nodata	&	\nodata	&	\nodata	\\
G50.07+0.06	&	0	&	0	&	\nodata	&	\nodata	&	\nodata	&	\nodata	\\
G53.88-0.18	&	0	&	0	&	\nodata	&	\nodata	&	\nodata	&	\nodata	\\
G75.75+0.75	&	0	&	0	&	\nodata	&	\nodata	&	\nodata	&	\nodata	\\
G76.38+0.63	&	0	&	0	&	\nodata	&	\nodata	&	\nodata	&	\nodata	\\
\enddata 
\end{deluxetable}

\begin{deluxetable}{lrrrcccrccc} 
\rotate
\tablecolumns{13} 
\tablewidth{0pc} 
\tablecaption{Observational Parameters at Center of MSX Core for CS J = 2 $\rightarrow$ 1 and CS J = 3 $\rightarrow$ 2} 
\tablehead{ 
&&&\multicolumn{4}{c}{CS J = 2 $\rightarrow$ 1} &\multicolumn{4}{c}{CS J = 3 $\rightarrow$ 2}\\
\tableline
\colhead{Core} & \colhead{$\Delta\alpha$}   & \colhead{$\Delta\delta$ }    & \colhead{$T_a^*$} & 
\colhead{$\Delta v$}    & \colhead{$v_{lsr}$}   & \colhead{$\int T_a^*dv$}    & \colhead{$T_a^*$} & \colhead{$\Delta v$}    & \colhead{$v_{lsr}$}   & \colhead{$\int T_a^*dv$ } \\
& \colhead{($''$)} & \colhead{($''$)} & \colhead{(K)} & \colhead{(km s$^{-1}$)} & \colhead{(km s$^{-1}$)} & \colhead{(K km s$^{-1}$)} & \colhead{(K)} & \colhead{(km s$^{-1}$)} & \colhead{(km s$^{-1}$)} & \colhead{(K km s$^{-1}$)} \\
\colhead{(1)} & \colhead{(2)} & \colhead{(3)} & \colhead{(4)} & \colhead{(5)} & \colhead{(6)} & \colhead{(7)} & \colhead{(8)} & \colhead{(9)} & \colhead{(10)}  & \colhead{(11)}
}

\startdata 
G05.85-0.23	&	25	&	-25	&	0.21	&	2.20	&	16.90	&	0.47	&	\nodata	&	\nodata	&	\nodata	&	\nodata	\\
G06.26-0.51	&	0	&	0	&	0.37	&	4.28	&	23.13	&	1.70	&	0.45	&	4.33	&	22.43	&	2.06	\\
G09.16+0.06	&	0	&	0	&	0.33	&	1.13	&	31.18	&	0.40	&	0.32	&	1.76	&	31.08	&	0.59	\\
G09.21-0.22	&	0	&	0	&	0.42	&	2.70	&	42.70	&	1.21	&	0.58	&	2.43	&	43.00	&	1.50	\\
G09.28-0.15	&	0	&	0	&	0.47	&	2.70	&	41.33	&	1.36	&	0.36	&	2.72	&	41.59	&	1.03	\\
G09.86-0.04	&	0	&	0	&	0.62	&	2.37	&	17.80	&	1.56	&	0.73	&	2.21	&	17.83	&	1.73	\\
G09.88-0.11	&	0	&	0	&	0.30	&	1.80	&	17.30	&	0.54	&	\nodata	&	\nodata	&	\nodata	&	\nodata	\\
G10.59-0.31	&	0	&	0	&	$<$ 0.20	&	\nodata	&	\nodata	&	\nodata	&	\nodata	&	\nodata	&	\nodata	&	\nodata	\\
G10.70-0.33 	&	0	&	0	&	$<$ 0.11	&	\nodata	&	\nodata	&	\nodata	&	\nodata	&	\nodata	&	\nodata	&	\nodata	\\
G10.99-0.09 	&	0	&	0	&	0.22	&	4.15	&	29.22	&	0.99	&	0.34	&	2.68	&	29.70	&	0.97	\\
G12.22+0.14	&	0	&	0	&	0.75	&	2.26	&	39.66	&	1.80	&	0.96	&	2.17	&	40.04	&	2.21	\\
G12.50-0.22	&	0	&	-25	&	0.44	&	2.64	&	35.97	&	1.23	&	0.48	&	1.95	&	35.99	&	0.99	\\
G14.33-0.57	&	0	&	0	&	0.51	&	2.34	&	19.58	&	1.26	&	\nodata	&	\nodata	&	\nodata	&	\nodata	\\
G19.28-0.39	&	0	&	0	&	0.24	&	1.10	&	54.00	&	0.26	&	0.10	&	1.04	&	53.24	&	0.22	\\
G19.37-0.03	&	0	&	0	&	0.73	&	3.63	&	26.97	&	2.82	&	1.36	&	4.20	&	26.82	&	6.05	\\
G19.40-0.01	&	0	&	0	&	0.27	&	3.05	&	26.49	&	0.87	&	\nodata	&	\nodata	&	\nodata	&	\nodata	\\
G23.37-0.29	&	0	&	0	&	0.44	&	4.67	&	77.88	&	2.20	&	\nodata	&	\nodata	&	\nodata	&	\nodata	\\
G23.48-0.53	&	0	&	0	&	0.17	&	4.18	&	63.78	&	0.23	&	$<$ 0.16	&	\nodata	&	\nodata	&	\nodata	\\
G24.05-0.22	&	25	&	0	&	0.26	&	2.70	&	81.67	&	0.75	&	0.40	&	2.74	&	81.57	&	1.17	\\
G24.16+0.08	&	0	&	0	&	$<$ 0.10	&	\nodata	&	\nodata	&	\nodata	&	$<$ 0.09	&	\nodata	& \nodata	& \nodata	\\
G25.99-0.06	&	25	&	25	&	0.35	&	2.20	&	90.18	&	0.82	&	0.41	&	2.85	&	89.78	&	1.26	\\
G30.14-0.07	&	0	&	0	&	$<$ 0.11	&	\nodata	&	\nodata	&	\nodata	&	\nodata	&	\nodata	&	\nodata	&	\nodata	\\
G30.49-0.39	&	0	&	0	&	$<$ 0.10	&	\nodata	&	\nodata	&	\nodata	&	$<$ 0.07	&	\nodata	&	\nodata	&	\nodata	\\
G30.53-0.27	&	0	&	0	&	0.24	&	7.30	&	102.90	&	1.73	&	\nodata	&	\nodata	&	\nodata	&	\nodata	\\
G30.89+0.14	&	0	&	0	&	0.15	&	0.74	&	95.76	&	0.39	&	\nodata	&	\nodata	&	\nodata	&	\nodata	\\
G30.98-0.15	&	-25	&	0	&	0.46	&	3.58	&	77.77	&	1.74	&	0.46	&	3.10	&	77.33	&	1.50	\\
G31.02-0.12	&	-25	&	0	&	0.21	&	3.29	&	76.73	&	0.72	&	0.16	&	3.30	&	77.16	&	0.78	\\
G32.01+0.05	&	0	&	0	&	0.50	&	5.55	&	95.83	&	2.94	&	0.47	&	5.37	&	95.49	&	2.68	\\
G33.82-0.22	&	0	&	0	&	0.45	&	0.95	&	11.51	&	0.45	&	\nodata	&	\nodata	&	\nodata	&	\nodata	\\
G34.63-1.03	&	0	&	0	&	$<$ 0.09	&	\nodata	&	\nodata	&	\nodata	&	$<$ 0.08	&	\nodata	&	\nodata	&	\nodata	\\
G34.74-0.12	&	-25	&	0	&	0.28	&	3.78	&	78.67	&	1.13	&	0.24	&	3.16	&	79.23	&	0.80	\\
G34.78-0.80	&	0	&	0	&	0.43	&	3.58	&	43.43	&	1.64	&	0.65	&	3.02	&	43.35	&	2.08	\\
G35.20-0.72	&	0	&	0	&	0.52	&	3.26	&	33.21	&	1.82	&	\nodata	&	\nodata	&	\nodata	&	\nodata	\\
G37.44+0.14	&	0	&	0	&	0.40	&	1.12	&	40.03	&	0.48	&	0.29	&	1.82	&	40.11	&	0.55	\\
G37.89-0.15	&	0	&	-25	&	0.57	&	0.66	&	12.96	&	0.40	&	0.31	&	1.00	&	12.95	&	0.32	\\
G43.64-0.82	&	0	&	0	&	$<$ 0.10	&	\nodata	&	\nodata	&	\nodata	&	$<$ 0.06	&	\nodata	&	\nodata	&	\nodata	\\
G43.78+0.05	&	0	&	0	&	$<$ 0.10	&	\nodata	&	\nodata	&	\nodata	&	$<$ 0.05	&	\nodata	&	\nodata	&	\nodata	\\
G50.07+0.06	&	0	&	0	&	$<$ 0.12	&	\nodata	&	\nodata	&	\nodata	&	0.54	&	1.63	&	54.66	&	0.05	\\
G53.88-0.18	&	0	&	0	&	$<$ 0.13	&	\nodata	&	\nodata	&	\nodata	&	$<$ 0.08	&	\nodata	&	\nodata	&	\nodata	\\
G75.75+0.75	&	0	&	0	&	$<$ 0.08	&	\nodata	&	\nodata	&	\nodata	&	$<$ 0.05	&	\nodata	&	\nodata	&	\nodata	\\
G76.38+0.63	&	0	&	0	&	$<$ 0.09	&	\nodata	&	\nodata	&	\nodata	&	$<$ 0.07	&	\nodata	&	\nodata	&	\nodata	\\
\enddata 
\end{deluxetable} 

\begin{deluxetable}{lrrrcccrccc} 
\rotate
\tablecolumns{13} 
\tablewidth{0pc} 
\tablecaption{Observational Parameters at Center of MSX Core for CS J = 5 $\rightarrow$ 4 and CS J = 7 $\rightarrow$ 6} 
\tablehead{ 
&&&\multicolumn{4}{c}{CS J = 5 $\rightarrow$ 4} &\multicolumn{4}{c}{CS J = 7 $\rightarrow$ 6}\\
\tableline
\colhead{Core} & \colhead{$\Delta\alpha$}   & \colhead{$\Delta\delta$ }    & \colhead{$T_a^*$} & 
\colhead{$\Delta v$}    & \colhead{$v_{lsr}$}   & \colhead{$\int T_a^*dv$}    & \colhead{$T_a^*$} & \colhead{$\Delta v$}    & \colhead{$v_{lsr}$}   & \colhead{$\int T_a^*dv$ } \\
& \colhead{($''$)} & \colhead{($''$)} & \colhead{(K)} & \colhead{(km s$^{-1}$)} & \colhead{(km s$^{-1}$)} & \colhead{(K km s$^{-1}$)} & \colhead{(K)} & \colhead{(km s$^{-1}$)} & \colhead{(km s$^{-1}$)} & \colhead{(K km s$^{-1}$)} \\
\colhead{(1)} & \colhead{(2)} & \colhead{(3)} & \colhead{(4)} & \colhead{(5)} & \colhead{(6)} & \colhead{(7)} & \colhead{(8)} & \colhead{(9)} & \colhead{(10)}  & \colhead{(11)}
}

\startdata 
G05.85-0.23	&	25	&	-25	&	\nodata	&	\nodata	&	\nodata	&	\nodata	&	\nodata	&	\nodata	&	\nodata	&	\nodata	\\
G06.26-0.51	&	0	&	0	&	0.13	&	4.07	&	19.91	&	0.56	&	\nodata	&	\nodata	&	\nodata	&	\nodata	\\
G09.16+0.06	&	0	&	0	&	\nodata	&	\nodata	&	\nodata	&	\nodata	&	\nodata	&	\nodata	&	\nodata	&	\nodata	\\
G09.21-0.22	&	0	&	0	&	0.08	&	2.45	&	41.73	&	0.53	&	\nodata	&	\nodata	&	\nodata	&	\nodata	\\
G09.28-0.15	&	0	&	0	&	0.22	&	2.60	&	41.66	&	0.23	&	\nodata	&	\nodata	&	\nodata	&	\nodata	\\
G09.86-0.04	&	0	&	0	&	0.12	&	1.50	&	17.60	&	0.19	&	\nodata	&	\nodata	&	\nodata	&	\nodata	\\
G09.88-0.11	&	0	&	0	&	\nodata	&	\nodata	&	\nodata	&	\nodata	&	\nodata	&	\nodata	&	\nodata	&	\nodata	\\
G10.59-0.31	&	0	&	0	&	\nodata	&	\nodata	&	\nodata	&	\nodata	&	\nodata	&	\nodata	&	\nodata	&	\nodata	\\
G10.70-0.33 	&	0	&	0	&	\nodata	&	\nodata	&	\nodata	&	\nodata	&	\nodata	&	\nodata	&	\nodata	&	\nodata	\\
G10.99-0.09 	&	0	&	0	&	0.07	&	3.20	&	30.31	&	0.08	&	0.47	&	3.20	&	29.66	&	0.83	\\
G12.22+0.14	&	0	&	0	&	0.25	&	2.20	&	39.60	&	1.06	&	\nodata	&	\nodata	&	\nodata	&	\nodata	\\
G12.50-0.22	&	0	&	-25	&	\nodata	&	\nodata	&	\nodata	&	\nodata &	0.46	&	2.30	&	37.27	&	2.91	\\
G14.33-0.57	&	0	&	0	&	\nodata	&	\nodata	&	\nodata	&	\nodata	&	\nodata	&	\nodata	&	\nodata	&	\nodata	\\
G19.28-0.39	&	0	&	0	&	\nodata	&	\nodata	&	\nodata	&	\nodata	&	\nodata	&	\nodata	&	\nodata	&	\nodata	\\
G19.37-0.03	&	0	&	0	&	0.28	&	3.80	&	27.14	&	0.57	&	\nodata	&	\nodata	&	\nodata	&	\nodata	\\
G19.40-0.01	&	0	&	0	&	\nodata	&	\nodata	&	\nodata	&	\nodata	&	\nodata	&	\nodata	&	\nodata	&	\nodata	\\
G23.37-0.29	&	0	&	0	&	\nodata	&	\nodata	&	\nodata	&	\nodata	&	\nodata	&	\nodata	&	\nodata	&	\nodata	\\
G23.48-0.53	&	0	&	0	&	\nodata	&	\nodata	&	\nodata	&	\nodata	&	\nodata	&	\nodata	&	\nodata	&	\nodata	\\
G24.05-0.22	&	25	&	0	&	\nodata	&	\nodata	&	\nodata	&	\nodata	&	0.24	&	3.30	&	80.23	&	3.12	\\
G24.16+0.08	&	0	&	0	&	\nodata	&	\nodata	&	\nodata	&	\nodata	&	\nodata	&	\nodata	&	\nodata	&	\nodata	\\
G25.99-0.06	&	25	&	25	&	0.26	&	2.50	&	89.24	&	1.96	&	\nodata	&	\nodata	&	\nodata	&	\nodata	\\
G30.14-0.07	&	0	&	0	&	\nodata	&	\nodata	&	\nodata	&	\nodata	&	\nodata	&	\nodata	&	\nodata	&	\nodata	\\
G30.49-0.39	&	0	&	0	&	\nodata	&	\nodata	&	\nodata	&	\nodata	&	\nodata	&	\nodata	&	\nodata	&	\nodata	\\
G30.53-0.27	&	0	&	0	&	\nodata	&	\nodata	&	\nodata	&	\nodata	&	\nodata	&	\nodata	&	\nodata	&	\nodata	\\
G30.89+0.14	&	0	&	0	&	\nodata	&	\nodata	&	\nodata	&	\nodata	&	\nodata	&	\nodata	&	\nodata	&	\nodata	\\
G30.98-0.15	&	-50	&	0	&	0.26	&	3.20	&	78.02	&	1.32	&	\nodata	&	\nodata	&	\nodata	&	\nodata	\\
G31.02-0.12	&	-25	&	0	&	\nodata	&	\nodata	&	\nodata	&	\nodata	&	\nodata	&	\nodata	&	\nodata	&	\nodata	\\
G32.01+0.05	&	0	&	0	&	0.20	&	5.40	&	95.96	&	0.46	&	\nodata	&	\nodata	&	\nodata	&	\nodata	\\
G33.82-0.22	&	0	&	0	&	\nodata	&	\nodata	&	\nodata	&	\nodata	&	\nodata	&	\nodata	&	\nodata	&	\nodata	\\
G34.63-1.03	&	0	&	0	&	\nodata	&	\nodata	&	\nodata	&	\nodata	&	\nodata	&	\nodata	&	\nodata	&	\nodata	\\
G34.74-0.12	&	-25	&	0	&	\nodata	&	\nodata	&	\nodata	&	\nodata	&	\nodata	&	\nodata	&	\nodata	&	\nodata	\\
G34.78-0.80	&	0	&	0	&	\nodata	&	\nodata	&	\nodata	&	\nodata	&	\nodata	&	\nodata	&	\nodata	&	\nodata	\\
G35.20-0.72	&	0	&	0	&	\nodata	&	\nodata	&	\nodata	&	\nodata	&	\nodata	&	\nodata	&	\nodata	&	\nodata	\\
G37.44+0.14	&	0	&	0	&	\nodata	&	\nodata	&	\nodata	&	\nodata	&	\nodata	&	\nodata	&	\nodata	&	\nodata	\\
G37.89-0.15	&	0	&	-25	&	\nodata	&	\nodata	&	\nodata	&	\nodata	&	\nodata	&	\nodata	&	\nodata	&	\nodata	\\
G43.64-0.82	&	0	&	0	&	\nodata	&	\nodata	&	\nodata	&	\nodata	&	\nodata	&	\nodata	&	\nodata	&	\nodata	\\
G43.78+0.05	&	0	&	0	&	\nodata	&	\nodata	&	\nodata	&	\nodata	&	\nodata	&	\nodata	&	\nodata	&	\nodata	\\
G50.07+0.06	&	0	&	0	&	\nodata	&	\nodata	&	\nodata	&	\nodata	&	\nodata	&	\nodata	&	\nodata	&	\nodata	\\
G53.88-0.18	&	0	&	0	&	\nodata	&	\nodata	&	\nodata	&	\nodata	&	\nodata	&	\nodata	&	\nodata	&	\nodata	\\
G75.75+0.75	&	0	&	0	&	\nodata	&	\nodata	&	\nodata	&	\nodata	&	\nodata	&	\nodata	&	\nodata	&	\nodata	\\
G76.38+0.63	&	0	&	0	&	\nodata	&	\nodata	&	\nodata	&	\nodata	&	\nodata	&	\nodata	&	\nodata	&	\nodata	\\
\enddata 
\end{deluxetable} 

\begin{deluxetable}{lccccl} 
\tablecolumns{6} 
\tablewidth{0pc} 
\tablecaption{Physical Conditions at Central Positions} 
\tablehead{ 
\colhead{Core} & \colhead{Category}  & \colhead{$N_{CO}$} & \colhead{$N_{N_2H^+}$} & \colhead{$n_{H_2}$} &\colhead{$N_{CS}$} \\
& & \colhead{(10$^{18}$ $cm^{-2}$)} & \colhead{(10$^{12}$ $cm^{-2}$)}  & \colhead{(10$^{5}$ $cm^{-3}$)} & \colhead{(10$^{12}$ $cm^{-2}$)}\\
\colhead{(1)} & \colhead{(2)} & \colhead{(3)} & \colhead{(4)} & \colhead{(5)} & \colhead{(6)}
}
\startdata 
G05.85-0.23	&	3	&	0.77	&	1.28	&	\nodata	&	4.55	*	\\
G06.26-0.51	&	1	&	0.88	&	\nodata	&	9.55	&	15.8		\\
G09.16+0.06	&	1	&	0.75	&	\nodata	&	6.03	&	3.80		\\
G09.21-0.22	&	2	&	\nodata	&	3.80	&	5.75	&	12.6		\\
G09.28-0.15	&	2	&	\nodata	&	3.90	&	12.9	&	10.0		\\
G09.86-0.04	&	2	&	\nodata	&	0.92	&	4.68	&	16.2		\\
G10.99-0.09 	&	2	&	1.22	&	4.46	&	6.92	&	10.0		\\
G12.22+0.14	&	2	&	\nodata	&	3.92	&	10.2	&	18.2		\\
G12.50-0.22	&	2	&	0.86	&	3.12	&	9.33	&	10.0		\\
G14.33-0.57	&	1	&	1.17	&	\nodata	&	\nodata	&	9.33		\\
G19.37-0.03	&	2	&	\nodata	&	3.74	&	29.5	&	44.7		\\
G19.40-0.01	&	3$^a$	&	\nodata	&	1.03	&	\nodata	&	8.41	*	\\
G23.37-0.29	&	2	&	2.04	&	3.06	&	\nodata	&	21.3	*	\\
G23.48-0.53	&	3	&	\nodata	&	1.56	&	\nodata	&	2.22	*	\\
G24.05-0.22	&	2	&	1.23	&	2.97	&	63.1	&	11.0		\\
G24.16+0.08	&	1	&	0.88	&	\nodata	&	\nodata	&	$<$1.84	**	\\
G25.99-0.06	&	2	&	\nodata	&	0.86	&	22.4	&	8.91		\\
G30.89+0.14	&	3	&	\nodata	&	1.34	&	\nodata	&	3.78	*	\\
G30.98-0.15	&	2	&	\nodata	&	4.82	&	15.8	&	13.8		\\
G31.02-0.12	&	3	&	0.67	&	1.68	&	3.24	&	6.03		\\
G32.01+0.05	&	2	&	2.37	&	7.30	&	10.5	&	23.4		\\
G34.63-1.03	&	2	&	0.31	&	1.22	&	\nodata	&	$<$9.58	**	\\
G34.74-0.12	&	2	&	1.86	&	2.04	&	4.37	&	10.7		\\
G34.78-0.80	&	2	&	1.19	&	0.35	&	1.55	&	18.6		\\
G35.20-0.72	&	2	&	1.09	&	3.22	&	\nodata	&	17.6	*	\\
G37.44+0.14	&	1	&	0.60	&	\nodata	&	2.95	&	4.68		\\
G37.89-0.15	&	2	&	0.34	&	0.48	&	1.20	&	4.79		\\
\enddata 

\tablecomments{Category 4 cores have not been included in this table.  A `\nodata' in columns 3 and 4 indicates no detectable emission for the given molecular line.   A '\nodata' in column 5 indicates that LVG analysis could not be performed and therefore no density was calculated.}
\tablenotetext{a}{ - 2 cores within the field of view, central core was identified as a category 3, however the 2nd object can be identified as a category 2}
\tablenotetext{*}{ - Did not have enough transitions of CS for LVG analysis, CS  J = 2 $\rightarrow$ 1 assumed to be in LTE and optically thin}
\tablenotetext{**}{ - Calculated using an upper limit to the intensity, but no detection above noise was made}

\end{deluxetable}

\begin{deluxetable}{lc} 
\tablecolumns{2} 
\tablewidth{0pc} 
\tablecaption{Chemical Abundances from Model} 
\tablehead{ 
\colhead{Species} & \colhead{Abundance}
}
\startdata 
e$^-$	&	1.46 x 10$^{-4}$	\\
N	&	4.50 x 10$^{-5}$	\\
O	&	3.52 x 10$^{-4}$	\\
C$^+$	&	1.46 x 10$^{-4}$	\\
S$^+$	&	4.00 x 10$^{-8}$	\\
Fe$^+$	&	6.00 x 10$^{-9}$	\\
He$^+$	&	2.53 x 10$^{-9}$	\\
Mg$^+$	&	4.00 x 10$^{-9}$	\\
Na$^+$	&	4.00 x 10$^{-9}$	\\
Si$^+$	&	4.00 x 10$^{-8}$	\\
H$_2$	&	1.00	\\
\enddata 

\tablecomments{All abundances are relative to H$_2$}

\end{deluxetable}

\begin{deluxetable}{lccccc} 
\tablecolumns{6} 
\tablewidth{0pc} 
\tablecaption{Table of Masses} 
\tablehead{ 
\colhead{Core} & \colhead{Distance}   & \colhead{Radius}    & \colhead{M$_{Jeans}$} & 
\colhead{M$_{Virial}$} & \colhead{M$_{Molecular}$} \\
 & \colhead{(kpc)} & \colhead{(pc)} & \colhead{(M$_{\odot}$)} & \colhead{(10$^3$ M$_{\odot}$)} & \colhead{(10$^3$ M$_{\odot}$)}\\
\colhead{(1)} & \colhead{(2)} & \colhead{(3)} & \colhead{(4)} & \colhead{(5)} & \colhead{(6)}
}
\startdata 
G05.85-0.23 & 3.14 & 0.38 & \nodata & 0.39 &  0.51 \\ 
G06.26-0.51 & 3.78 & 0.92 &  6 & 3.46 &  6.20 \\ 
G09.16+0.06 & 3.81 & 1.73 &  7 & 0.76 &  3.30 \\ 
G09.21-0.22 & 4.57 & 0.83 &  8 & 1.12 &  1.40 \\ 
G09.28-0.15 & 4.48 & 1.09 &  5 & 1.64 &  3.40 \\ 
G09.86-0.04 & 2.36 & 0.79 &  8 & 0.68 &  1.50 \\ 
G10.99-0.09 & 3.32 & 1.51 &  7 & 3.48 &  6.30 \\ 
G12.22+0.14 & 3.75 & 0.68 &  6 & 0.70 &  0.32 \\ 
G12.50-0.22 & 3.55 & 1.08 &  6 & 1.20 &  7.64 \\ 
G14.33-0.57 & 2.05 & 0.31 &  7 & 0.36 &  0.83 \\ 
G19.37-0.03 & 2.26 & 0.27 &  3 & 0.85 &  0.28 \\ 
G19.40-0.01 & 2.23 & 0.27 &  \nodata & 0.53 &  2.10 \\ 
G23.37-0.29 & 4.70 & 1.42 &  \nodata & 6.52 &  7.40 \\ 
G23.48-0.53 & 4.10 & 0.50 &  \nodata & 1.84 &  2.70 \\ 
G24.05-0.22 & 4.82 & 1.31 &  2 & 2.34 &  2.50 \\ 
G24.16+0.08	&	3.46	&	0.42	&	\nodata	&	\nodata	&	2.60	\\
G25.99-0.06 & 5.15 & 0.78 &  4 & 1.04 &  0.68 \\ 
G30.89+0.14 & 5.65 & 2.05 &  \nodata & 0.24 & 11.00 \\ 
G30.98-0.15 & 4.63 & 0.56 &  5 & 1.28 &  1.90 \\ 
G31.02-0.12 & 4.56 & 1.11 & 10 & 2.54 &  7.30 \\ 
G32.01+0.05 & 5.77 & 0.52 &  6 & 3.24 & 21.70 \\ 
G34.63-1.03	&	0.84	&	0.10	&	\nodata	&	\nodata	&	0.12	\\
G34.74-0.12 & 4.86 & 0.74 &  9 & 1.88 &  3.27 \\ 
G34.78-0.80 & 2.80 & 1.19 & 15 & 2.73 &  2.90 \\ 
G35.20-0.72 & 2.17 & 0.26 &  \nodata & 0.58 &  2.50 \\ 
G37.44+0.14 & 2.60 & 0.39 & 11 & 0.18 &  0.09 \\ 
G37.89-0.15 & 0.82 & 0.30 & 17 & 0.04 &  0.06 \\ 
\enddata 

\tablecomments{ '\nodata' indicates that LVG analysis could not be performed. Therefore an H$_2$ density was not available to calculate the Jeans mass.}

\end{deluxetable} 


\begin{references}
Aikawa, Y., Ohashi, N., Inutsuka, S., Herbst, E., \& Takakuwa, S. 2001, ApJ, 552, 639	\\
Ballesteros-Pareded, J.  2006, MNRAS, 372, 443 \\
Bergin, E. A., Langer, W. D., \& Goldsmith, P. F. 1995, ApJ, 441, 222	\\
Bergin, E. A. \& Tafalla, M. 2007  \araa , 45, 339-396. \\ 
Beuther, H., Walsh, A., Schilke, P., Sridharan, T. K., Menten, K. M., \& Wyrowski, F.  2002, \aap 390, 289-298. \\
Beuther, H., Schilke, P., Menten, K. M., Motte, F., Sridharan, T. K., \& Wyrowski, F. 2002, ApJ, 566, 945 \\
Beuther, H. \& Sridharan, T. K. 2007 \apj , 668, 348-358. \\
Beuther, H. \& Steinacker, J.  2007 \apjl ,656, L85-L88. \\
Bonnell, I. A., \& Davies, M. B. 1998, ApJ, 295, 691	\\
Butler, M. J. \& Tan, J. C.  2009 \apj , 696, 484-497 \\
Carey, S. J., Clark, F. O., Egan, M. P., Price, S. D., Shipman, R. F., \& Kuckar, T. A. 1998,  ApJ, 508, 721	\\
Caselli, P., Benson, P. J., Myers, P. C., \& Tafalla, M. 2002, ApJ, 572, 238	\\
Chambers, E. T., Jackson, J. M., Rathborne, J. M., \& Simon, R.  2009, \apjs , 181, 360-390. \\
Churchwell, E. 2002, ARA\&A, 40, 27	\\
Crapsi, A., Caselli, P., Walmsley, C. M., Myers, P. C., Tafalla, M., Lee, C. W., \& Bourke, T. L. 2005, ApJ, 619, 379 	\\
Dopita, M. 1991, PASA, 9, 234 \\
Du, F., \&  Yang, J.  2008,  \apj , 686, 384-398 \\
Fich, M., Blitz, L., \& Stark, A. 1989, ApJ, 342, 272 \\
Garay, G.,  Fa$\acute{u}$ndez, S., Mardones, D., Bronfman, L., Chini, R., \& Nyman, L. 2004, ApJ, 610, 313	\\
Garay, G. \& Lizano, S. 1999 \pasp , 111, 1049-1087.\\ 
Hasegawa, T. I., \& Herbst, E. 1993, MNRAS, 261, 83	\\
Indebetouw, R., et al. 2005, ApJ, 619, 931 \\
Knez, C., Shirley, Y. L., Evans, N. J., II, \& Mueller, K. E. 2002, ASPC, 267, 375	\\
Kutner, M. L., \& Ulich, B. L. 1981, ApJ, 250, 341	\\
Lada, C. J. \& Lada, E. A.  2003  \araa , 41, 57-115. \\
Lada, E. A., Strom, K. M., \& Myers, P. C. 1993, PRPL, 245	\\
Lada, E. A., Evans, N. J., II, \& Falgarone, E.  1997, ApJ, 488, 286	\\
Lee, J.E., Bergin, E.A., \& Evans, N.J. II 2004, ApJ, 617, 360 \\
Le Teuff, Y. H., Millar, T. J., \& Markwick, A. J. 2000, A\&A, 146, 157	\\
Leurini, S., Schilke, P., Wyrowski, F., \& Menten, K. M. 2007, A\&A, 466, 215 \\
McKee, C. F. 1986, Ap\&SS, 118, 383\\
McKee, C. F., \& Tan, J. C. 2003, ApJ, 585, 850	\\
Menten, K. M., Pillai, T., \& and Wyrowski, F. 2005, in Massive Star Birth: A Crossroads of Astrophysics, IAU 227, 23-34 \\ 
Molinari, S., Brand, J., Cesaroni, R., \& Palla, F. 1996, A\&A, 308, 573\\
Mueller, K. E., Shirley, Y. L., Evans, N. J., II, \& Jacobson, H. R.  2002, ApJS, 143, 469	\\
Nakano, T., \& Umebayashi, T. 1986, MNRAS, 221, 319	\\
Pillai, T., Wyrowski, F., Carey, S. J., \& Menten K. M. 2006, A\&A, 450, 569 \\
Pillai, T., Wyrowski, F., Hatchell, J., Gibb, A. G., \& Thompson, M. A. 2007, A\&A, 467, 207 \\
Plume, R., Jaffe, D. T., \& Evans, N. J., II 1992, ApJS, 78, 505	\\
Plume, R., Jaffe, D. T., Evans, N. J., II, Martin-Pintado, J., \& Gomez-Gonzalez, J. 1997, ApJ, 476, 730	\\
Price, S. D., et al. 1996, AAS, 28, 1341	\\
Ragan, S. E., Bergin, E. A., Plume, R., Gibson, D. L., Wilner, D. J., O'Brien, S., \& Hails, E. 2006, ApJS, 166, 567	\\
Ragan, S.E., Bergin, E.A., \& Gutermuth, R.A. 2009, ApJ, 698, 324 \\
Rathborne, J.M., Jackson, J.M., \& Simon, R. 2006,  ApJ, 641, 389 \\ 
Rathborne, J. M., Jackson, J. M., Zhang, Q., \& Simon, R. 2008, ApJ, 689, 1141 \\
Redman, R. O., Feldman, P. A., C$\hat{o}$t$\acute{e}$, S., Wyrowski, F., Carey, S. J., \& Egan, M. P. 2002, ASPC, 267, 409\\
Rudolph, A. L., Bachiller, R., Rieu, N. Q., Van Trung, D., Palmer, P., \& Welch, W. J. 2001, ApJ, 558, 204	\\
Shirley, Y. L., Evans, N. J., II, Young, K. E., Knez, C., \& Jaffe,  D. T. 2003, ApJS, 149, 375	\\
Snell, R. L., Langer, W. D., \& Frerking, M. A. 1982, ApJ, 255, 149	\\
Sridharan, T. K., Beuther, H.,  Saito, M., Wyrowski, F., \& Shilke, P. 2005, ApJ, 634, 57	\\
Sridharan, T. K., Beuther, H.,  Shilke, P., Menten, K., \& Wyrowski, F. 2002, ApJ, 566, 931	\\
Tafalla, M. Myers, P. C., Caselli, P., \& Walmsley, C. M. 2004, Ap\&SS, 292, 347	\\
Wall, W. F. 2007, MNRAS, 379, 674 \\
Wang, Y., Zhang, Q., Rathborne, J. M., Jackson, J., \& Wu, Y. 2006, \apjl , 651, L125-L128 \\
Wu, Y., Zhang, Q., Chen, H., Yang, C., \& Ho, P. T. P. 2000, ASPC, 217, 98	\\
Zhou, S., Wu, Y., Evans, N. J., II, Fuller, G. A., \& Myers, P. C. 1989, ApJ, 346, 168	\\
\end{references}
\end{document}